# Exploring Solar-Terrestrial Interactions via Multiple Observers

## A White Paper for the Voyage 2050 long-term plan in the ESA Science Programme

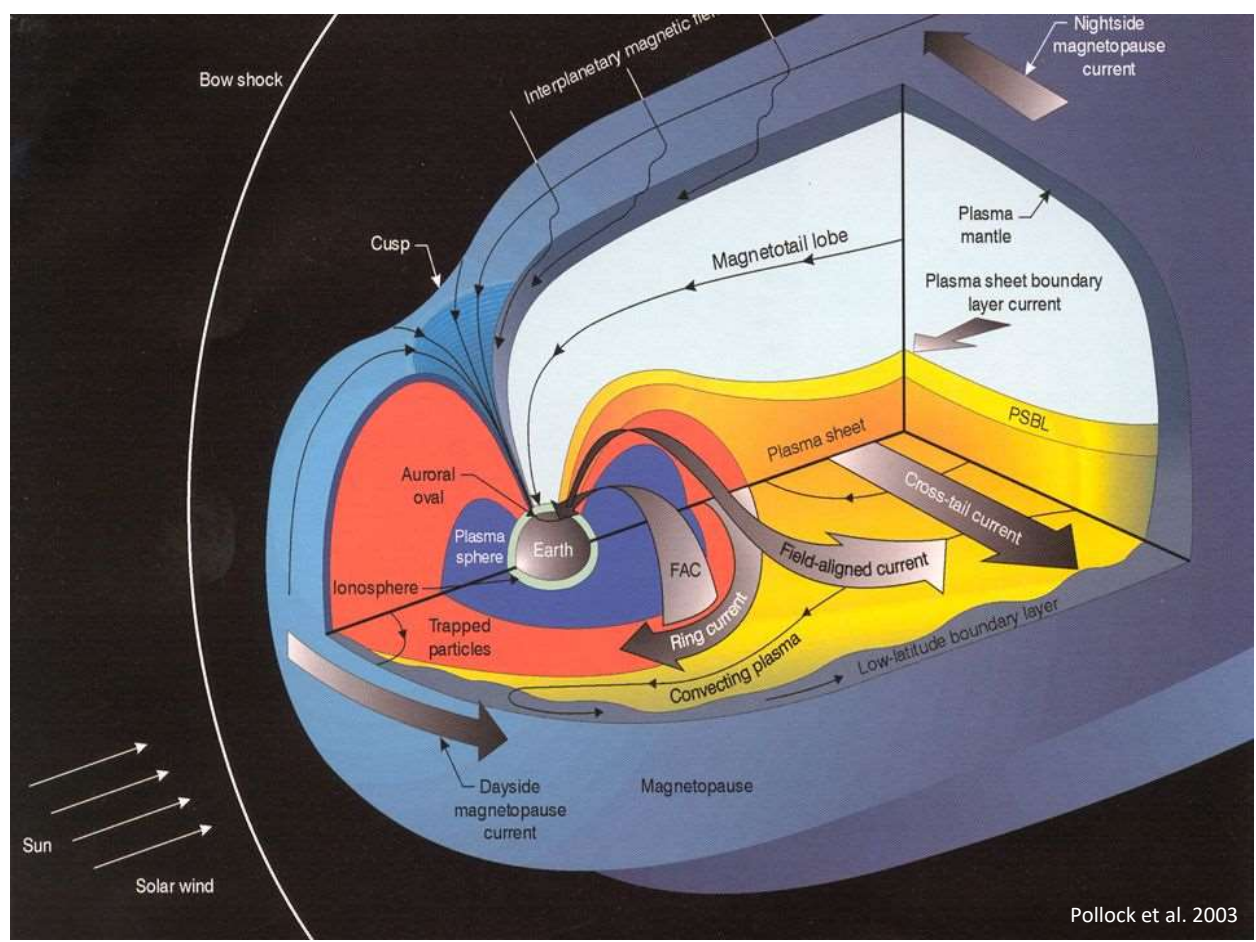

Pollock et al. 2003

**Contact Scientist**

**Graziella Branduardi-Raymont**

Mullard Space Science Laboratory, University College London
Holmbury St Mary, Dorking, Surrey RH5 6NT, UK
Tel. +44 (0)1483 204133
g.branduardi-raymont@ucl.ac.uk

# Exploring Solar-Terrestrial Interactions via Multiple Observers

## Executive summary

The central question we propose to address is: ***How does solar wind energy flow through the Earth's magnetosphere, how is it converted and distributed?*** This is a fundamental science question expressing our need to understand how the Sun creates the heliosphere, and how the planets interact with the solar wind and its magnetic field. This is not just matter of scientific curiosity – it also addresses a clear and pressing practical problem. As our world becomes ever more dependent on complex technology – both in space and on the ground – society becomes more exposed to the vagaries of space weather, the conditions on the Sun and in the solar wind, magnetosphere, ionosphere and thermosphere that can influence the performance and reliability of technological systems and endanger human life and health.

This fundamental question breaks down to several sub-questions: 1) How is energy transferred from the solar wind to the magnetosphere at the magnetopause? 2) What are the external and internal drivers of the different magnetospheric regimes? 3) How does energy circulate through the magnetotail? 4) How do behaviours in the North and South hemispheres relate to each other? 5) What are the sources and losses of ring current and radiation belt plasma in the inner magnetosphere? 6) How does feedback from the inner magnetosphere influence dayside and nightside processes?

Much knowledge has already been acquired through observations in space and on the ground over the past decades, but the infant stage of space weather forecasting demonstrates that we still have a vast amount of learning to do. We can tackle this issue in two ways: 1) By using multiple spacecraft (e.g. Cluster, THEMIS, Swarm, MMS) measuring conditions in situ in the magnetosphere in order to make sense of the fundamental small scale processes that enable transport and coupling, or 2) By taking a global approach to observations of the conditions that prevail throughout geospace in order to quantify the global effects of external drivers.

A global approach is now being taken by a number of space missions under development (e.g. SMILE, LEXI) and the first tantalising results of their exploration will be available in the next decade. Here we propose the next step-up in the quest for developing a complete understanding of how the Sun gives rise to and controls the Earth's plasma environment: a tomographic imaging approach which can be achieved with an M-class mission consisting of two spacecraft enabling global imaging of magnetopause and cusps, auroral regions, plasmasphere and ring current, alongside in situ measurements. Such a mission is going to be crucial on the way to achieve scientific closure on the question of solar-terrestrial interactions.

### 1. Science questions to be addressed

The Earth's magnetic field extends into space (see Fig. 1), where its interaction with the supersonic solar wind plasma leads to the formation of the magnetosphere. The solar wind flow compresses the sunward side of the magnetosphere but drags the nightside out into a long magnetotail. A collisionless bow shock stands upstream from the magnetopause in the supersonic solar wind. The shocked solar wind plasma flows around the magnetosphere through the magnetosheath.

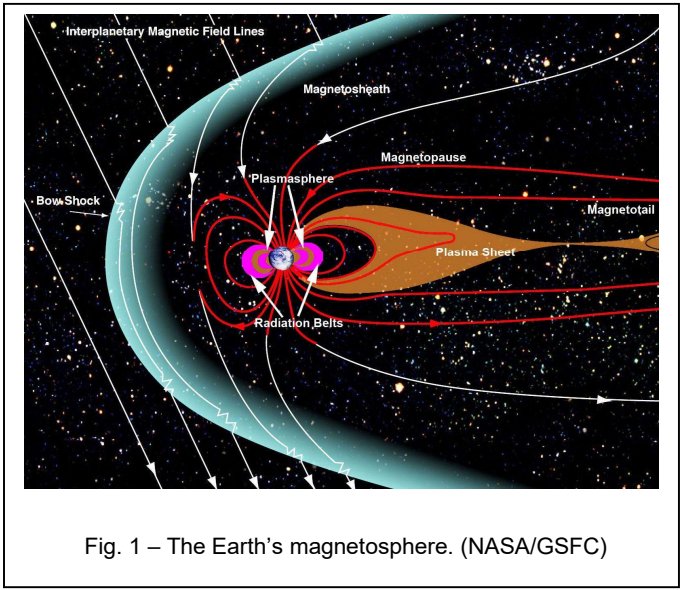

Fig. 1 – The Earth's magnetosphere. (NASA/GSFC)

A relatively sharp transition from dense, shocked, solar wind plasmas to tenuous magnetospheric plasmas marks the magnetopause. High latitude cusps denote locations where field lines divide to close either in the opposite hemisphere or far down the magnetotail. Open field lines within the cusps provide an opportunity for solar wind plasma to penetrate deep into the magnetosphere, all the way to the ionosphere. The position and shape of the magnetopause change constantly as the Earth's magnetosphere responds to varying solar wind dynamic pressures and interplanetary magnetic field orientations.

A basic component of magnetospheric activity is the Dungey cycle (Dungey 1961), illustrated in Fig. 2: intervals of dayside reconnection under southward interplanetary magnetic field (IMF) enable the magnetic flux content of the magnetotail lobes to increase and energy to be stored. Reconnection in the tail and magnetospheric convection bring plasma back to the dayside magnetosphere. The stored energy is intermittently and explosively released in *geomagnetic substorms*, and is associated with bright auroral displays in polar regions (Angelopoulos et al. 2008). *Geomagnetic storms*, usually driven by coronal mass ejections (CMEs) or corotating interaction regions (CIRs), and associated with prolonged periods of strong southward IMF, represent a severe space weather threat with the greatest capacity to disrupt everyday life throughout the world (e.g. affecting satellite subsystems and astronaut wellbeing in space, telecommunications, electrical infrastructures, and pipelines on the ground).

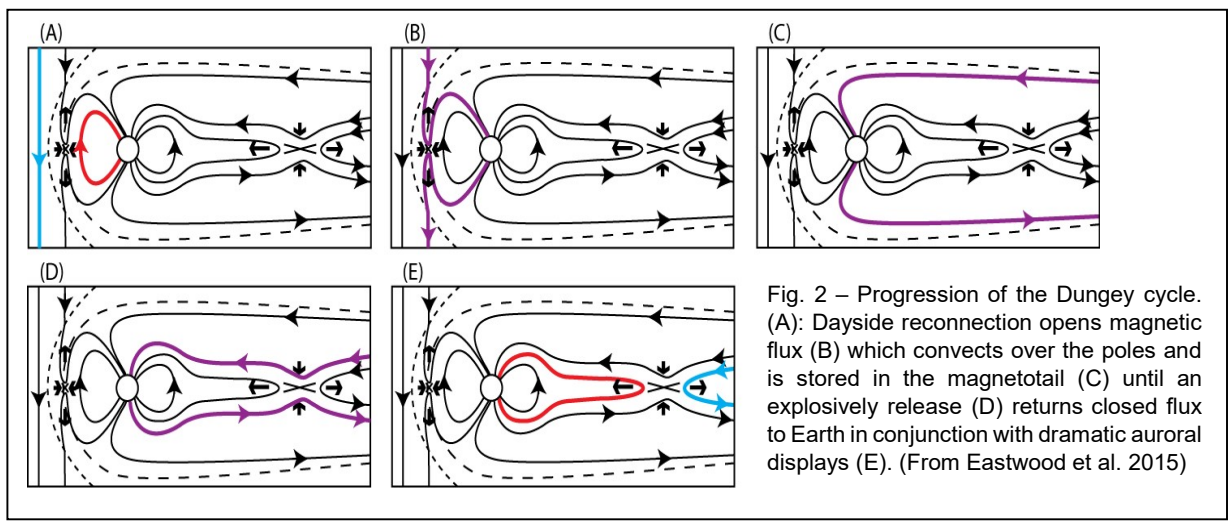

Fig. 2 – Progression of the Dungey cycle. (A): Dayside reconnection opens magnetic flux (B) which convects over the poles and is stored in the magnetotail (C) until an explosively release (D) returns closed flux to Earth in conjunction with dramatic auroral displays (E). (From Eastwood et al. 2015)

The energy input from the solar wind is continuously converted from the dynamic to the electromagnetic form and back, while plasma is moving through the magnetospheric and ionospheric regions (Fig. 2). The solar wind dynamic energy is partly converted to electromagnetic energy at the bow shock. Magnetic reconnection at the magnetopause plays a crucial role in the energy flux penetration from the solar wind into the magnetosphere, and this represents an opposite energy conversion process from the electromagnetic to the dynamic form. Due to the dayside magnetic reconnection, the magnetic energy is transmitted to the nightside magnetosphere, accumulated in the magnetotail and then suddenly released by nightside reconnection. The energetic particles from the magnetotail move towards the Earth, form the ring current and partly precipitate in the auroral regions. The precipitated particles and magnetospheric-ionospheric currents deliver the solar wind energy into the ionosphere. Ionospheric outflow constitutes an opposite mass and energy stream from the ionosphere into the magnetosphere, which influences nightside and dayside reconnection rates.

*Magnetospheric modes*

Depending on solar wind driving, the magnetosphere can change between different dynamic regimes, or magnetospheric modes, e.g. stationary magnetospheric convection, saw-tooth oscillations, isolated substorms and sequences of substorms and storms (e.g. DeJong et al. 2009; Pulkkinen et al. 2010, Sergeev et al. 1996; Walach et al. 2017, Hubert et al. 2017).

Transitions between the modes may result from changes in the upstream solar wind conditions and/or be consequences of the internal magnetospheric-ionospheric dynamics. In the first case, the IMF as well as solar wind plasma parameters regulate the energy input into the magnetosphere. The direction of the IMF plays an important role in the coupling between the solar wind and the magnetosphere. For example, during northward IMF conditions, coupling is complex, consisting of lobe reconnection as well as plasma transfer due to Kelvin-Helmholtz waves at the magnetospheric flanks (Taylor et al. 2008, Otto and Fairfield 2000). Quasi-radial IMF results in the foreshock formation in the dayside region, high-speed jets in the magnetosheath and magnetopause deformation (Plaschke et al. 2018). However, magnetic activity in the magnetosphere is usually stronger during southward IMF, when the IMF merges with the Earth's magnetic field at the dayside magnetopause. On the other hand, magnetospheric modes may be consequence of internal magnetospheric – ionospheric dynamics. For example, Brambles et al. (2011) reported that the ionospheric O+ outflows can generate saw-tooth oscillations. While the different modes of magnetospheric dynamics are recognised, the questions of what external and internal conditions drive a particular mode and under what conditions mode transformations occur are still open. These are very significant issues in space weather research and have strong bearing on future successful forecast of the magnetospheric state and dynamics. Such issues are especially important as humans increasingly depend on technological infrastructures which can be adversely affected by space weather.

Global simulations, in situ and remote measurements have provided evidence for different reconnection modes at the magnetopause. Reconnection can be steady (Sonnerup et al. 1981) or bursty (Russell and Elphic 1978). The extension of this process across the dayside and flank magnetopause is also unclear. Understanding the temporal and spatial properties of magnetopause reconnection (e.g. variable solar wind driving, temporal behaviour of the reconnection process, length of reconnection line) is essential as they define how much energy is transferred from the solar wind into the magnetosphere. The energy return flow from the tail towards inner magnetosphere and ionosphere depends on the properties of substorms. Magnetospheric substorms may be triggered by solar wind variations and by either southward or northward IMF turnings, albeit with different efficiencies (e.g. Liou et al. 2003, Wild et al. 2009). On the other hand, some substorms occur without apparent IMF perturbations (Hsu and McPherron 2004).

Magnetospheric storms can be considered as additional, extreme, states of the Earth's magnetospheric dynamics, which are often the consequence of extreme heliospheric drivers, CMEs and CIRs (e.g. Denton et al. 2006). They are expected to occur on average at least once a month (McPherron1995). A storm is characterised by radiation belts build-up (although also depletion in some cases, Turner et al. 2013) and development of a strong ring current; however, it can also be accompanied by a number of other phenomena, e.g. by a series of substorms, by the motion of the auroral oval equatorward, strong and complex field-aligned currents and high fluxes of energetic particles. The magnetospheric dynamics during storm time is a subject of space weather research, and forecast of such dynamics is crucial for protecting satellites, ground infrastructure and human health. It is now understood that the strongest storms are driven by CME/CME or CME/CIR interactions (Liu et al. 2015). However, internal magnetospheric processes are less clear, especially the role of the substorms in the development of the storm time ring current is still under active debate (e.g. Runge et al. 2018, Keller et al. 2005).

**All the above illustrates that we have acquired basic, qualitative knowledge about magnetospheric dynamics, e.g. about southward IMF triggering reconnection at the magnetopause, subsequent reconnection in the tail and how this affects the ring current during storms. However, we do not know how to quantify the energy circulation in the magnetosphere, we do not have precise knowledge of when and how the system changes as a result of given inputs, hence we are currently unable to predict strength and temporal variations of storms and substorms.**

*Magnetopause and cusps*

The solar wind propagates through the bow shock and magnetosheath and interacts with the dayside magnetopause. During southward IMF intervals, the IMF reconnects with the magnetospheric magnetic field, and solar wind energy enters the magnetosphere. This results in reconfiguration of the magnetospheric – ionospheric (MI) currents, and the magnetopause in the subsolar region shifts Earthward (both empirical and numerical MHD models confirm this Earthward motion). The ground PC index also indicates when IMF discontinuities reach the subsolar magnetopause and the merging electric field at the magnetopause changes (e.g. Troshichev et al., 2006; Samsonov et al., 2018). Moreover, southward IMF and strong solar wind dynamic pressure applied to the dayside magnetopause cause global intensification of the aurora which can be observed for example at FUV wavelengths (e.g. Boudouridis et al. 2007). This almost immediate response to solar wind dynamic pressure variations is not connected with the energy accumulation in the magnetotail, instead energy and mass come directly from the dayside magnetopause. Energetic ions delivered through reconnection at the magnetopause increase the energetic ion population in the dayside and near-Earth magnetosphere (Luo et al. 2017).

Coupling of the solar wind with the magnetosphere also takes place through the magnetospheric cusps, where solar wind particles travelling along magnetic field lines can penetrate directly into the magnetosphere. The peculiar magnetic topology of the cusps means that they play a pivotal role in magnetospheric dynamics: they are the sole locations where solar wind has direct access to low altitudes (e.g. Cargill et al. 2005). During magnetopause reconnection, solar wind energy, mass and momentum are transferred through the cusps into the magnetosphere. The equatorward boundary of the cusp region is often identified as the boundary between closed dayside field lines and open ones. For southward IMF the cusp lies on the open field lines because of dayside magnetopause reconnection. The amount of open flux depends on the rate of reconnection both on the dayside magnetopause and in the nightside magnetotail plasma sheet. Therefore, the cusps move to lower and higher latitudes as the open field line region expands and contracts, respectively. It is thus of key importance to continuously measure how the cusps respond to northward and southward turnings of the IMF, since this is intimately related to the strength of the solar wind – magnetosphere coupling. Since the cusps are the endpoints of a large portion of the magnetospheric magnetic field, their structure, azimuthal extent, local time and latitudinal location give information about a larger context than any other structure within the magnetosphere (see Sibeck et al. 2018 for a review). Since component and antiparallel reconnection models predict different cusp locations, by studying the cusps we gather deeper insight into magnetopause reconnection.

**Although multipoint in situ measurements of the cusps have provided evidence for the proposed reconnection mechanisms (e.g. stationary or temporal, component or antiparallel), global imaging can distinguish between them, quantify their significance and also establish any asymmetry between North and South cusps and ionospheric regions.**

*Magnetotail reconnection and aurora formation*

Magnetotail reconnection begins when a significant magnetic energy has been stored in the magnetotail. The amount of the accumulated energy can be estimated by the size of the polar cap (Shukhtina et al. 2005). The area of open flux within the polar cap changes directly in response to the amount of open flux in the magnetotail lobes, which are defined by the dayside and magnetotail reconnection processes. Auroral activity and variations of the ground magnetospheric indices (e.g. SuperMAG indices, such as SME, SMU, SML; for acronyms see Newell and Gjerloev 2011) are the signatures of the subsequent energy release in the ionosphere and characterise substorm dynamics through variations of the MI currents. As field aligned currents provide coupling between the magnetosphere and ionosphere, the variability of the MI currents describes substorm dynamics.

The THEMIS mission was aimed at answering the key questions of magnetotail substorm timing and whether reconnection at the near-Earth X-line is initiated first and drives the current disruption closer to the Earth, or the substorm starts with current disruption at ~10 – 12 $R_E$ and this launches the reconnection down the tail. However, results from THEMIS have shown that the magnetotail processes are more complex and the questions posed have not been answered in full.

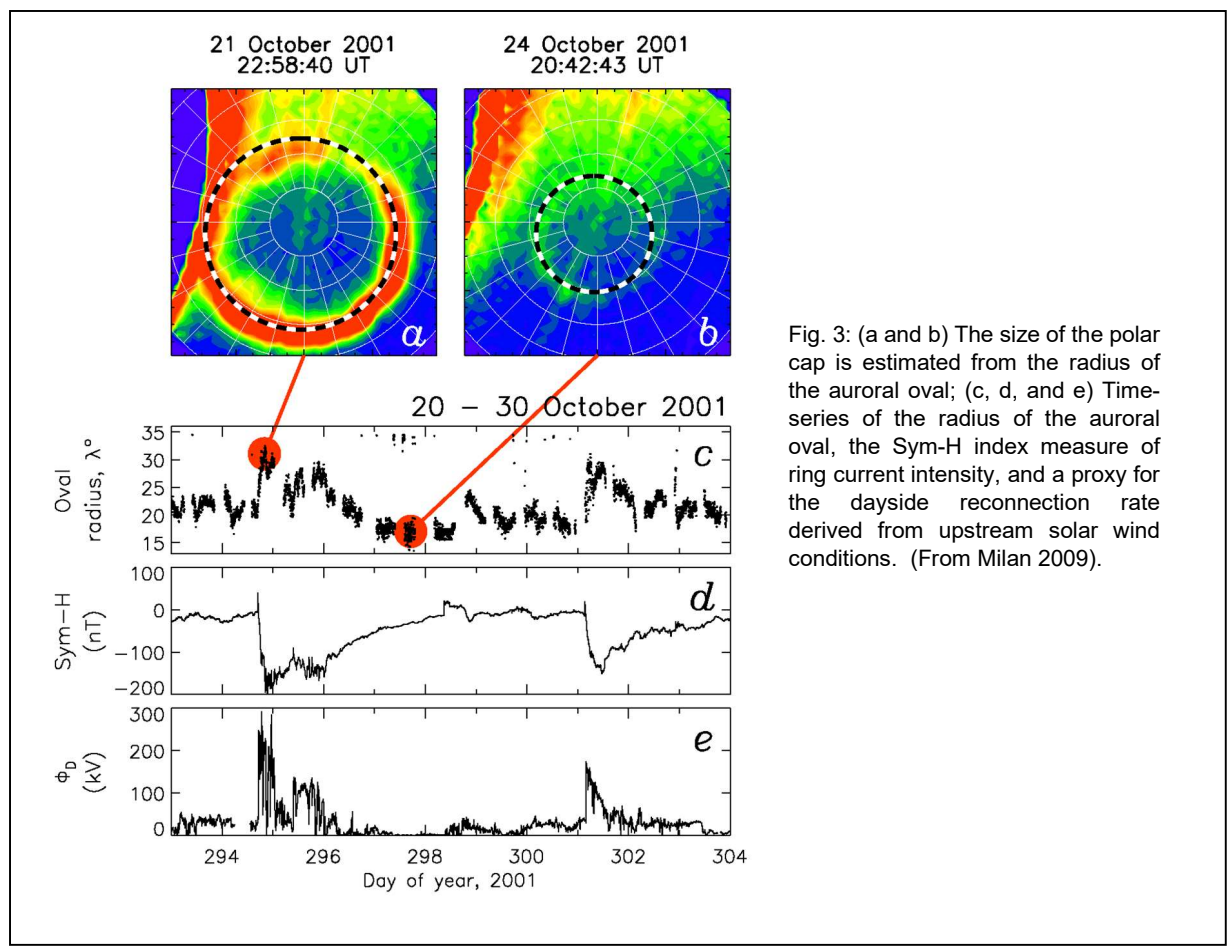

Fig. 3: (a and b) The size of the polar cap is estimated from the radius of the auroral oval; (c, d, and e) Time-series of the radius of the auroral oval, the Sym-H index measure of ring current intensity, and a proxy for the dayside reconnection rate derived from upstream solar wind conditions. (From Milan 2009).

Global simulations, in situ and remote measurements provide evidence for a wealth of magnetotail reconnection modes (e.g. McPherron et al. 2008; Walach et al. 2017; Dejong et al. 2009). For example, isolated substorms (e.g. Akasofu 2017) require a period of energy storage which can be initiated by southward IMF turnings. These isolated tail modes exhibit a ~1 hr growth phase corresponding to an expansion of the auroral oval leading to a transient onset and auroral brightening. This is followed by a ~30 min expansion phase characterised by dynamic aurora moving poleward and stretching ~3 hours of Local Time across the night side sky, and by strong electric fields in the inner nightside magnetosphere. The expansion phase is followed by a ~1 hour recovery phase during which the aurora dims.

Fig. 3 shows FUV imaging of the Northern polar cap by IMAGE. An estimate of the size of the polar cap can be derived from the radius of the auroral oval. These measurements have necessarily been non-continuous in the past due to the orbits of auroral imaging missions (data gaps in panel c), hindering progress in fully understanding solar wind – magnetosphere coupling. Short time-scale variations in polar cap size correspond to substorms, but large discontinuities exist over some data gaps indicating that the storm behaviour is only partially captured. This is an area where the continuous (over 40 hours) auroral monitoring by SMILE for the first time will make unbroken determination of the rates of magnetic reconnection and the factors that influence these. However, SMILE cannot do conjugate monitoring of the aurorae to capture asymmetries, which is particularly necessary under northward IMF and under IMF with strong dusk – dawn component.

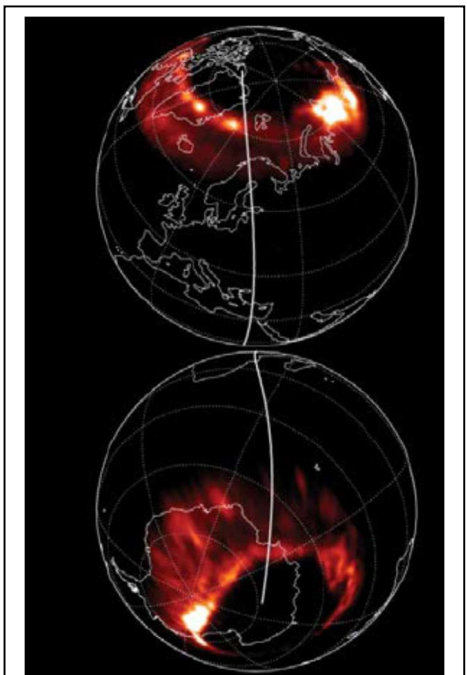

Fig. 4 – Simultaneous views of the North and South aurorae obtained by Polar and IMAGE. (From Laundal & Ostgaard 2009)

When the IMF is directed northwards, lobe reconnection takes place at high latitudes, tailwards of the cusps openings. Under sufficiently dense solar wind conditions, the footprint of the reconnection site is visible as an auroral 'spot' within the noon-sector of the polar cap (Milan et al. 2000, Frey et al. 2002). Unlike southward IMF reconnection, lobe reconnection is not constrained to occur equally in the northern and southern hemispheres and indeed the interrelation of lobe reconnection in the two hemispheres is unknown. The cadence of auroral imagers flown in the past has not been sufficient to properly analyse the dynamics of cusp spots. Consequently the conjugate nature of these cusp and substorm features is completely unknown as simultaneous observations of both hemispheres are extremely rare, due to the lack of coordination between past missions. Fig. 4 illustrates simultaneous views of the North and South aurorae taken by the Polar and IMAGE spacecraft: some features are symmetrical, but many are not, challenging the current understanding of magnetic field mapping between hemispheres and our understanding of the fundamental processes leading to auroral emission. These may be the first clear observations of interhemispheric currents due to seasonal differences (Richmond and Roble 1987, Benkevich et al. 2000).

**The very important but poorly understood issue of asymmetries in the magnetosphere – ionosphere current system has not been adequately tackled so far. Significant open questions remain about the physics of substorm complete development, including its initialisation process. Such questions could be answered for the first time by continuous and conjugate FUV monitoring, with identical instrumentation, of both auroral ovals.**

*Ring current sources and losses*

Energetic particles injected by the magnetotail reconnection contribute to the formation of the ring current. Energetic positively charged ions can undergo charge exchange with local neutrals turning into energetic neutral atoms (ENA), whose direction, energy and species composition can be measured. By tracking the intensity of the created ENAs by charge exchange from the ring current as a function of time during isolated substorms, saw-tooth events, and storm-time substorms it is possible to quantify how individual events and prolonged intervals of nightside reconnection contribute to ring current plasma intensities, its sources and losses. ENA imaging (e.g. Mitchell et al. 2001, see Fig. 5) determines the depth to which ion injections penetrate, their azimuthal extent, the degree to which ions are energised, spectral slopes, and composition. ENA directly measures loss via charge exchange as a function of species (H, O), energy, location, and time. The second main contributor to the ring current formation is a large-scale electric field during stationary convective intervals in the nightside magnetosphere. The relative roles of the large-scale electric field and substorms in the buildup of the ring current are a subject of active debate.

Ring current ions with energies of 30-300 keV at distances of 4 to 5 $R_E$ on average from Earth originate in the Earth's plasma sheet. Each magnetotail reconnection mode supplies ions to the ring current. However, ring current intensities do not increase indefinitely. During storms, ring current intensities initially intensify, then decay, first rapidly over ~8 hour during the early recovery phase, and then more slowly over the next several days (Hamilton et al. 1988 – See Fig. 5).

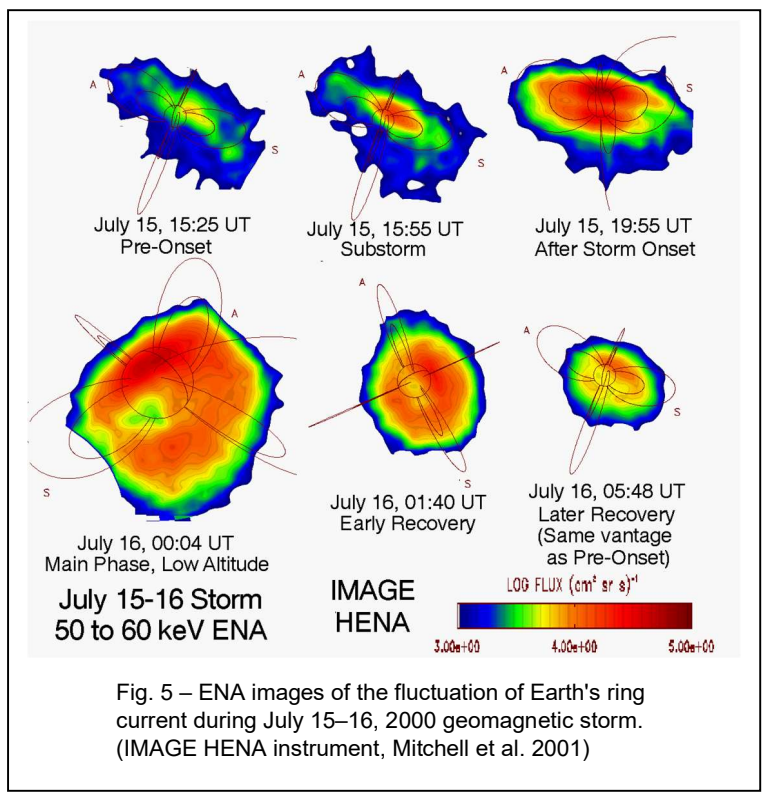

Fig. 5 – ENA images of the fluctuation of Earth's ring current during July 15–16, 2000 geomagnetic storm. (IMAGE HENA instrument, Mitchell et al. 2001)

Proposed mechanisms for ring current loss, in addition to charge exchange, include wave-particle induced precipitation and magnetopause outflow. Daglis et al. (1999) concluded that charge exchange with the exosphere is the main mechanism for ring current decay. As the convection electric field decays during the late recovery phase the slower charge exchange loss dominates ring current decay on completely closed drift paths (Takahashi et al. 1990, Liemohn et al. 2001). EMIC wave-particle interactions and precipitation may become important during the main phase of geomagnetic storms (e.g. Gonzalez et al. 1989). Significant dropouts of relativistic electrons can also take place as consequence of magnetopause 'shadowing', following compression of the magnetopause, e.g. by increase in solar wind dynamic pressure, and subsequent loss of trapped particles while drifting around the Earth (Herrera et al. 2016).

We have also no knowledge of the 3-D shape of the ring current, which is crucial in the 3-D reconstruction of the magnetic field via accurate modelling (e.g. Tsyganenko 2013).

Tomographic measurements by ENA imagers mounted on at least two spacecraft on appropriately chosen orbits for the first time would provide the missing data.

**Open questions are: 1) How efficiently the magnetotail response modes described above energise the ring current ions, 2) How transport and loss mechanisms affect the subsequent evolution of the ring current, 3) What is the relative importance of electric fields and substorms in building up the ring current, and of charge exchange, wave-particle interactions and magnetopause shadowing in ring current losses. Determination of the 3-D structure of the ring current would help improve significantly our magnetic field models.**

*Plasmasphere feedback*

The plasmasphere is a region of cold plasma of ionospheric origin which is trapped within the co-rotating portion of the inner magnetosphere. Its distribution has significant modifying effects on particle-particle and wave-particle interactions taking place in the inner magnetosphere. EUV emissions from the plasmasphere (see Fig. 6) helps to interpret the feedback from the inner magnetosphere to the dayside magnetopause.

The size of the plasmasphere depends on the balance between co-rotation and convection, the latter depending on the strength of the solar wind interaction and reconnection in the magnetotail so that the plasmasphere expands during quiet conditions and contracts when the convection is strongly driven, e.g. during geomagnetic storms.

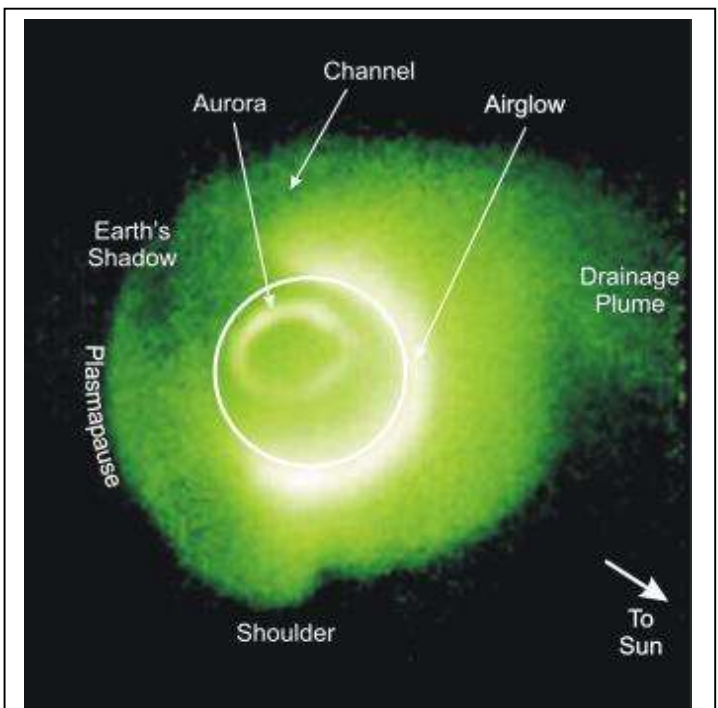

Fig. 6 – Earth's plasmasphere and plume as measured by IMAGE's EUV instrument. (From Sandel et al. 2003)

The inner magnetosphere is not simply the final stop in the circulation of energy in the coupled solar wind – magnetosphere system. Rather, the inner magnetosphere hosts dynamic processes that have important feedback effects on the solar wind – magnetosphere interaction.

Plasmaspheric plumes frequently become entrained in magnetopause reconnection (McFadden et al. 2008) particularly during geomagnetic storms when plumes persist for days (Borovsky et al. 2007). Since reconnection rates are inversely proportional to plasma densities (Cassak and Shay 2007), the arrival of a high density plasmaspheric plume at the magnetopause quenches reconnection.

**How the plasmasphere is refilled and how the plasma distributes itself along the field lines are still open questions, as well as the time-dependent processes that lead to plasmaspheric loss and to the transport of cold plasma within the magnetosphere.**

## 2. Observing methods to answer open science questions

*Magnetopause and cusps imaging*

Solar wind charge exchange is now recognised as the atomic process generating soft X-rays (0.2 – 2.0 keV, a range including a large number of K- and L-shell atomic transitions) in a variety of astrophysical scenarios, and in the Earth's magnetosheath and magnetospheric cusps too. This has enabled a novel approach to mapping the magnetosphere in a global way: wide field of view (FOV) X-ray images allow us to study kinetic physics in geospace on the global scale. So far applications of this technique (e.g. SMILE Soft X-ray Imager) have been limited to single imagers: these will return 2-D images from which the 3-D structure of the magnetopause needs to be reconstructed, with line of sight integration introducing uncertainties. These could be alleviated substantially by tomographic measurements performed by having imagers on at least two spacecraft on appropriately chosen orbits.

Soft X-ray images with high sensitivity and good spatial and temporal resolution are required to reach closure on some of the science questions stated above, e.g. Earthward magnetopause shifting for southward IMF; formation, dynamics and properties of Flux Transfer Events, or FTEs; magnetopause motion after solar wind directional discontinuities; high-speed jets downstream of the quasi-parallel bow shock; magnetopause indentations, and possibly Kelvin-Helmholtz vortices.

*Auroral imaging*

Auroral emissions are produced by the precipitation of energetic electrons and protons. High energy precipitating electrons excite atmospheric constituents by impact, with emissions in the FUV band from Lyman-Birge-Hopfield transitions of $N_2$ (160 – 180 nm) and OI at 135.6 nm. Electrons with energies in excess of 30 keV produce X-rays by bremsstrahlung. Proton aurorae are seen at the footprints of the magnetospheric cusps, where they are injected from the magnetosheath by lobe reconnection (Frey et al. 2002) although their relationship to similar electron aurorae (Milan et al. 2000) is still unknown. Proton precipitation produces Ly$\alpha$ emission as protons charge exchange with atmospheric constituents to create excited H atoms. In order to distinguish the emission of (down-travelling) precipitating protons from the geocoronal Ly$\alpha$ the redshifted wing of the Ly$\alpha$ line is imaged. This requires a system that efficiently rejects both the geocoronal Ly$\alpha$ emitted at 121.6 nm and the NI multiplet at 120 nm. IMAGE was able to image the proton aurora.

Breakthrough science can arise from operating two three-axis stabilised spacecraft, allowing continuous and conjugate imaging of both aurorae, with high cadence imaging and the capability to suppress dayglow contamination.

*Ring current imaging*

ENAs are produced when singly positively charged energetic ions undergo charge exchange collisions with cold neutral atoms or molecules. The ions become neutral and travel on unaffected by electromagnetic fields. In addition to carrying spectral and directional information of the energetic ions, the ENAs also provide direct measurement of their species composition. In the Earth's magnetosphere the ring current charge-exchanges with the geocorona at high altitudes, emitting ENAs that allow the ring current and plasma sheet ion populations to be imaged (Roelof et al. 2004).

ENA imaging maps the neutral hydrogen and oxygen ions generated when ring current and plasma sheet ions encounter exospheric neutrals, providing the information needed to track plasma sheet thinning and recovery, ring current growth and decay via substorm particle injections, precipitation, charge exchange, and magnetopause outflow. ENA imaging further identifies ring current effects on the magnetopause and nightside reconnection modes. Simultaneous monitoring of magnetopause and ring current allows to investigate directly and quantify magnetopause shadowing, i.e. the loss of energetic particles through the magnetopause when this is closer to the Earth, within the global loss of magnetospheric plasma.

*Plasmasphere imaging*

Global images of the plasmasphere can be obtained by observing EUV sunlight at 30.4 nm resonantly scattered from singly-ionised Helium, a minor magnetospheric constituent which allows extrapolation of overall magnetospheric density (Sandel et al. 2000). $He^+$ 30.4 nm is the brightest ion emission from the plasmasphere, it is spectrally isolated, and the background at this wavelength is negligible. Measurements can be easily interpreted because the plasmaspheric $He^+$ emission is optically thin, so its brightness is directly proportional to the $He^+$ column abundance.

EUV imaging can track the plasmapause to quantify the convection that occurs in response to dayside and nightside magnetotail reconnection modes, to identify locations where wave-particle interactions may drive ring current ion precipitation, and to determine when and where plasmaspheric plumes may affect dayside reconnection. EUV observations of plasmapause motion distinguish between and quantify steady and impulsive electric fields applied to the inner magnetosphere.

*In situ measurements*

The imaging observations outlined above require to be set into context by the availability of simultaneous in situ measurements. These could be gathered by monitoring spacecraft at the L1 point and propagated to near Earth orbit, although continuity and strict simultaneity cannot be assured. There are a few clear advantages of measuring immediately upstream as opposed to relying on L1. In situ monitoring of plasma conditions from locations outside the bow shock and in the Earth's vicinity, where remote imaging is performed, is a much more appropriate solution: especially if multiple spacecraft are involved, it quantifies the solar wind input and determines orientation and structure of solar wind discontinuities, arrival times at the magnetopause, and impact on the magnetosphere with much better accuracy than that available from propagating observations from any monitor at L1. If the spacecraft is in the foreshock then foreshock bubbles may be observed in situ and add context to the images (i.e. they could explain possible dawn-dusk asymmetries), which would otherwise be missed at L1. Also, interplanetary coronal mass ejections (ICMEs) are very dynamic and data at L1 do not take into account the expected evolution of the structures which may be important for triggering substorms.

Simultaneous in situ measurements in the northern and southern hemispheres (and on dawn and dusk flanks) can provide information about inclination of solar wind discontinuities and also about homogeneity of the solar wind stream (answering questions such as: Do we observe the same solar wind parameters on a separation of 40 to 60 $R_E$?). Very useful measurements could be made in the magnetotail (since probably there will be two 'tail seasons' per year). These would include magnetic field strength in the lobes (closely related to the expanding/contracting polar cap and solar wind pressure effects), and magnetic field and particle measurements in the plasma sheet at down-tail distances, where the near-Earth reconnection line is expected to be located.

*Possible complementary observing methods*

The key magnetospheric processes resulting in energy transfer and partition, and the transient structures which they create, involve accelerated and energised plasma, hence coverage (by imaging and in situ measurements) at high energies (e.g. tens of keV and above) is the way to identify and characterise these processes. Looking relatively far into the future, global measurement requirements expected to cover higher energy ranges could include high energy X-ray imagers and also tracers for the energetic particles themselves as part of the in situ package, in order to understand coupling between regions.

*Ground measurements*

Ground-based all-sky imagers and magnetometers distributed throughout Canada, Alaska and Northern Europe, and in Antarctica, are an essential complement to the space-borne instruments by providing the high time and spatial resolution observations needed to track the development of critical nightside auroral microstructures, including those that herald substorm onset. Global networks of ionospheric radars (e.g. SuperDARN) provide measurements of convection, which in turn quantifies reconnection rates (as well as the expanding/contracting polar cap). Combined space- and ground-based images (such as those from all-sky auroral cameras) place mesoscale auroral structures in their global context in a manner impossible during the IMAGE and THEMIS eras, and throughout almost the entire POLAR era.

### 3. Validation of magnetospheric models

An essential part of researching solar-terrestrial relationships is the validation of empirical and numerical magnetospheric models. Global magnetospheric conditions and dynamics can be simulated by means of numerical and empirical (and semi-empirical) models. A large number of models have already been developed for different magnetospheric regions (bow shock and magnetopause, magnetotail, inner magnetosphere, ionosphere and upper atmosphere, auroral regions). Some models are local and developed to simulate a particular region or for predicting some specific parameters, and some models are global and able to reproduce the global magnetospheric dynamics self-consistently, i.e. taking into account interrelations between different regions. The model development is important, because we can both better understand the physics of the processes and make predictions of extreme magnetospheric-ionospheric conditions which are important for space weather forecast.

International and national meteorological and space agencies have used some of these models already. In particular, for aurora forecast NOAA uses the OVATION-Prime model, which predicts global 2-D images of the auroral activity depending on solar wind conditions, while the University of Michigan's Geospace model can predict Kp and Dst indices as well as regional magnetic variations, and the Relativistic Electron Forecast Model (REFM) is used for high-energy electron fluence at geosynchronous orbit (see www.swpc.noaa.gov). The Geospace model is a part of the Space Weather Modeling Framework (SWMF) developed at the University of Michigan, simulates the full time-dependent 3D geospace environment (Earth's magnetosphere, ring current and ionosphere) and predicts global space weather parameters such as induced magnetic perturbations in space and on the Earth's surface. The current version of the Geospace model includes the global magnetospheric MHD model BATS-R-US and a kinetic model of the inner magnetosphere (Liemohn et al. 2018). The European space community has also developed several models, e.g. the TRANSPLANET ionospheric model (http://transplanet.irap.omp.eu/), and the same have done other space agencies, mostly for high-energetic particles and ionospheric disturbances, based predominantly on empirical models.

We anticipate that global magnetospheric models will become even more significant in the future because the magnetosphere is a region where processes in different parts are closely related to each other. One of the main approaches will be using combinations of fluid and kinetic models. These models can simulate both the propagation of the solar wind from the solar corona to the magnetosphere and its subsequent interaction with the magnetosphere and ionosphere, ultimately predicting geomagnetic disturbances and ionospheric conditions crucially important for human civilization and infrastructures. Another possible approach in the modelling is using artificial intelligence systems.

Irrespective of which one is used, any model has to be fully validated. However, this is not achievable at the present time. Most past and present space missions provide in situ measurements which cannot reflect the global magnetospheric configuration because the observations can be completely different when changing locations even slightly. Recent multi-spacecraft missions (Cluster, THEMIS, MMS) do not solve this problem completely because the observations are still sparse and local. Global context can be obtained by statistics, but this requires years of measurements with varying orbits (e.g. Cluster, THEMIS), and even then the coverage is limited for different solar wind conditions. The alternative of using many spacecraft is extremely expensive and impractical budget-wise for one agency. Imaging can produce global coverage almost immediately, for the immediate upstream conditions and at an affordable cost.

We only achieve some global observations near the Earth, e.g. using a large set of ground magnetometers and all-sky cameras; global magnetospheric imaging has been only sporadic so far. Using direct global measurements we can simultaneously validate both empirical models and combined MHD-kinetic models in several principal magnetospheric regions. Moreover, global imaging can underpin the development of a new generation of models which satisfy all observational constraints from the start.

Missions which use global imaging techniques have already been in operation (Polar, IMAGE) or are in preparation (SMILE, LEXI). However, to progress we need higher accuracy, better spatial resolution and simultaneous coverage of different magnetospheric regions by using different techniques in order to reconstruct the global magnetospheric dynamics and validate future magnetospheric models. We also need stereo vision, i.e. simultaneous observations from at least two different locations, in order to reconstruct 3-D distributions of magnetospheric parameters from 2-D global images. All magnetospheric boundaries and structures are three-dimensional and they are not necessarily approximated by simple mathematical functions as often assumed. Our current descriptions of the magnetopause and ring current shapes are examples of such simplified interpolations. Using stereo vision, we can reconstruct the real shape of this boundaries and current layers without any a priori assumptions.

### 4. How can a space mission address the open science questions?

Remote global magnetospheric images together with in situ solar wind measurements near the bow shock and available ground magnetospheric indices (which reflect the intensity of main magnetospheric current systems as observed by a large set of ground magnetometers) will provide complete information about the magnetospheric state, its variations and evolution. Using global magnetospheric imaging, we can collect simultaneously essential information about the positions and physical conditions of the main magnetospheric boundaries, i.e. the magnetopause, bow shock, magnetospheric cusps, the auroral oval, the ring current and plasmapause. Using this full array of measurements, we obtain a complete description of the magnetospheric state and can distinguish different magnetospheric modes. We also can specify how the global magnetospheric physical state and positions of magnetospheric boundaries change when the magnetosphere switches from one mode to another.

The science questions raised in section 1 have been addressed so far by a number of multipoint missions such as Cluster, THEMIS, Van Allen Probes and MMS, which target the substorm cycle, radiation belts transport and loss processes, the microphysics of reconnection and particle acceleration, and provide tantalising evidences for competing solar wind – magnetosphere interaction modes. However, they do not provide the global view necessary to distinguish between proposed interaction modes, determine their occurrence patterns and hence quantify their global significances.

In the near future the physics of solar-terrestrial interactions is going to be probed with dedicated space missions under development and planned (e.g. CuPID, launch 2020; LEXI, 2022; SMILE, 2023; see Sibeck et al. 2018). These apply the novel technique of soft X-ray imaging of the Earth's magnetosphere, usually coupled with well established FUV imaging of the auroral oval and in situ measurements. These missions will tackle some of the open questions, but will be unable to explore issues linking the solar wind – magnetosphere coupling with the ring current and plasmasphere, and especially those issues (e.g. energy transformation between different magnetospheric and ionospheric regions, taking into account asymmetries between northern and southern hemispheres) where the combined global view from multiple vantage points is required. Moreover, we can expect that the new knowledge provided by forthcoming missions will lead to new lines of investigation, requiring a step up in observing facilities in the longer term.

A comprehensive approach to studying solar-terrestrial interactions on the global scale could be achieved by a space mission capable of providing stereo vision for tomography studies of the Earth's magnetosphere: this would include X-ray imaging of the dayside magnetosheath and the magnetospheric cusps, coupled with simultaneous FUV monitoring of both North and South aurorae, imaging of the plasmasphere in the EUV and in ENA for the ring current. Using two spacecraft, 3-axis stabilised, in appropriately selected orbits, and two identical payloads, would allow significantly better reconstruction of the 3-D shape of the magnetopause, ring current and plasmasphere. Integral part of each payload would also be an in situ instrument package dedicated to providing self-sufficient monitoring of solar wind and magnetosheath plasma conditions.

**Orbit(s)**

In order to achieve a good vantage point while maintaining appropriate spatial resolution a large geocentric distance (e.g. 30 $R_E$) is suggested (a lower distance would be advantageous for auroral imaging at high resolution, but would hinder the tomography capability).

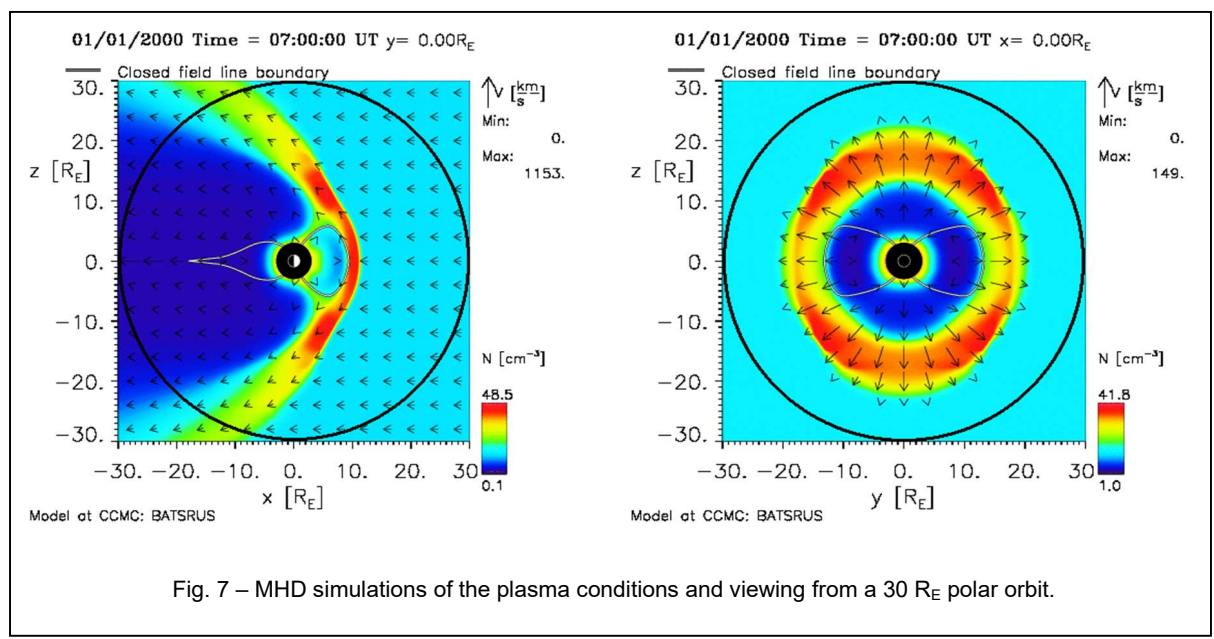

Fig. 7 – MHD simulations of the plasma conditions and viewing from a 30 $R_E$ polar orbit.

Two circular, highly inclined, polar orbits, with a 9.65 day period and phased 90° away from each other, would allow observations of magnetospheric plasma structures and the solar wind input from both polar and equatorial vantage points and achieve the goals of tomography and conjugate auroral monitoring discussed above. The 90° phasing, unlike 180°, allows simultaneous continuous monitoring of both aurorae, although only twice an orbit at nadir. Fig. 7 shows simulated side and polar views of the magnetosphere from a 30 $R_E$ orbit using the CCMC BATSRUS model.

A study by our NASA GSFC Core Proposing Team members has shown that a 30 $R_E$ circular polar orbit can be reached with lunar assist; two spacecraft, properly phased, can observe simultaneously both aurorae for long intervals over the 9.5 day orbital period, with the added bonus that circular orbits do not need adjusting. For some orbits, simultaneous observations on the dayside and nightside may also be possible, and in some cases crossings the bow shock from the solar wind into the magnetosheath, allowing us to measure different regimes.

**Payload**

X-ray imager

An enhanced version of the SMILE Soft X-ray Imager (see the ESA definition study report at https://www.cosmos.esa.int/documents/1655046/0/SMILE_RedBook_ESA_SCI_2018_1.pdf and Fig. 8) can be envisaged.

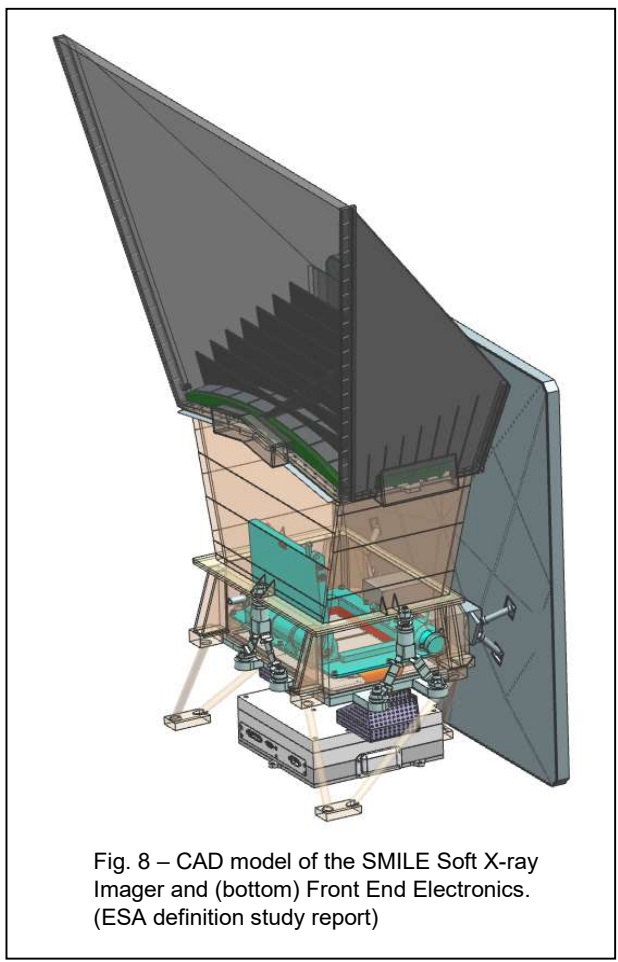

Fig. 8 – CAD model of the SMILE Soft X-ray Imager and (bottom) Front End Electronics. (ESA definition study report)

From a geocentric distance of 30 $R_E$ it is possible to capture the entire dayside magnetopause within a 30° x 30° FOV without sacrificing spatial resolution: Fig. 9 shows simulated soft X-ray images corresponding to observations made from two spacecraft with 90° phase shift and slightly different aim points in order to observe a larger area.

This is based on low mass lobster-eye optic technology, combined with large frame CCD or microchannel plate detecting devices. The former detectors provide energy resolution (~50 eV at 0.5 keV with current technology) but are susceptible to radiation damage, while the latter offer no spectral resolution, are less radiation sensitive and require high voltages.

Given the current detector developments for X-ray astrophysics missions, it is likely that the detector of choice for a future mission will be an active-pixel sensor using CMOS technology, offering all the performance benefits of CCDs with increased radiation hardness.

Auroral FUV imager

The auroral imager requirements can be derived from experience with previous missions (e.g. Polar, IMAGE). The new science emerges from the fact of having two spacecraft, permitting continuous and conjugate imaging, and spacecraft that are 3-axis stabilised, allowing continuous pointing and high cadence imaging with long exposure times.

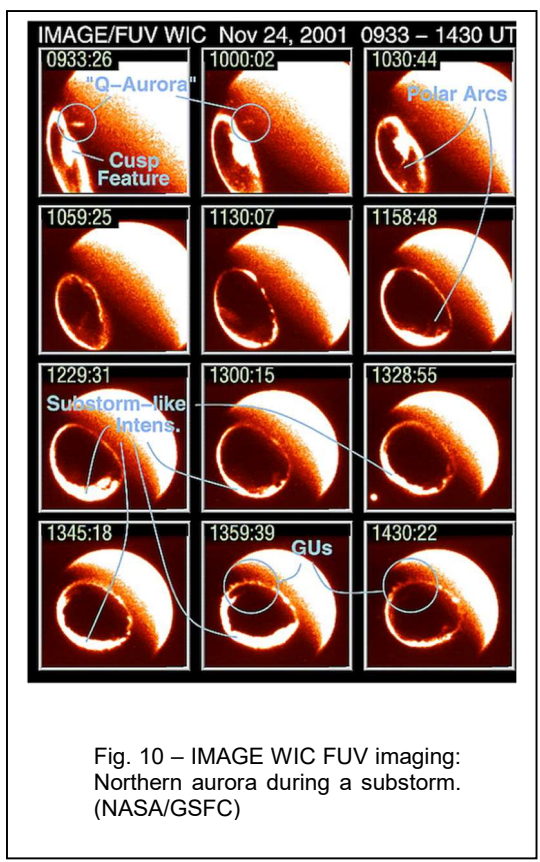

Fig. 10 – IMAGE WIC FUV imaging: Northern aurora during a substorm. (NASA/GSFC)

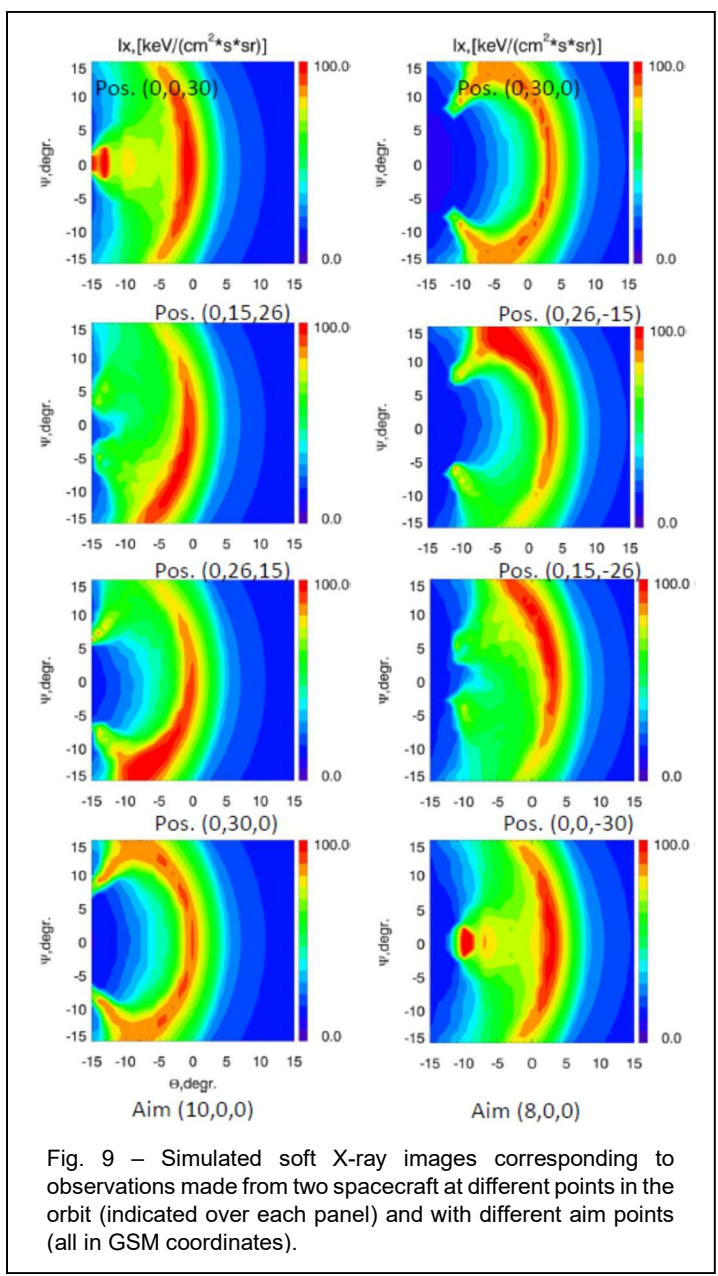

Fig. 9 – Simulated soft X-ray images corresponding to observations made from two spacecraft at different points in the orbit (indicated over each panel) and with different aim points (all in GSM coordinates).

The ability to suppress non-auroral emissions is vital for the science objectives, since the dayside aurora will be sunlit in at least one of the hemispheres most of the time (see Fig. 10 for an example from IMAGE WIC). Deposition of thin-film reflective coatings on mirrors with several reflecting surfaces, as well as on the detecting devices, enable isolation of the desired FUV passband for electron aurora (e.g. 160 – 180 nm), and provide many orders of magnitude visible light suppression. Alternative instrument designs can be considered for proton aurora imaging (at Ly$\alpha$, e.g. Mende et al. 2000). A 5° x 5° FOV is appropriate at a geocentric distance of 30 $R_E$.

The spatial resolution needs to be of the order of 100 km, similar to previous imaging missions, with a cadence of 2 min.

ENA imager

ENAs are generated in the terrestrial magnetosphere through charge exchange processes between magnetically trapped energetic ions and cold neutral gas. An ENA camera can record the arrival directions, energies and mass species of magnetospheric ENAs as well as indirectly provide global images of the spatial, energy and mass species distributions of their parent ion populations.

Fig. 11 shows a schematic of the IMAGE HENA instrument, with a FOV of 120° x 90°, covering the energy range 20 – 500 keV per nucleon, with energy resolution $\Delta E/E < 0.25$ and velocity resolution of 50 km s$^{-1}$. After an electrostatic deflection assembly at several keV positive potential has prevented charged particles from entering the camera, the remaining ENAs pass through a Time of Flight device which allows to compute their incoming direction and their velocity. The energy deposited in a solid state detector (SSD in Fig. 11) enables mass discrimination (separation of hydrogen and oxygen atoms).

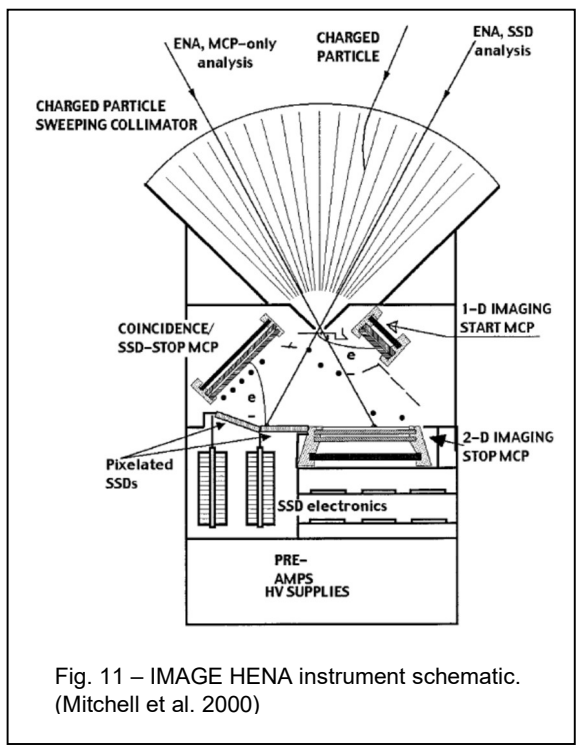

Fig. 11 – IMAGE HENA instrument schematic. (Mitchell et al. 2000)

A spatial resolution of a couple of $R_E$ and cadence of 30 min are adequate. Similar requirements apply to the study of the plasma sheet.

Plasmasphere EUV imager

The imaging requirements can be derived from heritage of the EUV instrument flown on IMAGE, which was spinning and carried three camera sensors for a total 84° x 30° FOV at an apogee of 7 $R_E$. From inspection of Fig. 6, where the white circle represents the size of the Earth, a 10 $R_E$ wide FOV, or 20° x 20° at 30 $R_E$ geocentric distance, is appropriate in order to provide, in a single exposure, a map of the entire plasmasphere from the outside, including plasmasphere plumes. A spatial resolution of ~0.3 $R_E$ and a 10 min cadence are required in order to satisfy science requirement.

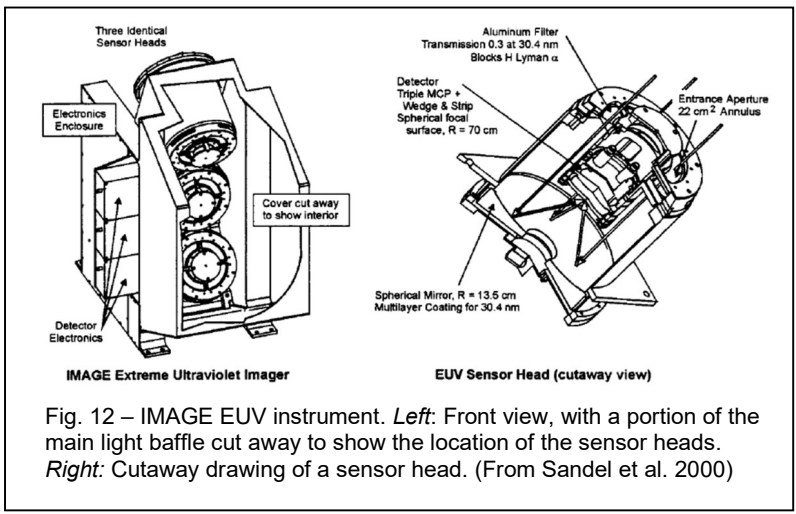

Fig. 12 – IMAGE EUV instrument. *Left*: Front view, with a portion of the main light baffle cut away to show the location of the sensor heads. *Right:* Cutaway drawing of a sensor head. (From Sandel et al. 2000)

Taking the design of IMAGE EUV (Sandel et al. 2000 – see Fig. 12) as an example, the entrance aperture of the imager incorporates an Aluminium filter that transmits the He$^+$ 30.4 nm line, while excluding the bright geocoronal H Ly$\alpha$ line at 121.6 nm. Light that goes through the filter reaches the mirror, which has a multilayer coating to provide good reflectivity at the target wavelength and low at 58.4 nm (this He I emission is expected to be weak in the plasmasphere, but can be quite bright in Earth's ionosphere). The mirror focuses the light on a sensor comprising bare microchannel plates (avoiding a photcathode that could increase the response to the contaminating H Ly$\alpha$ line) and a wedge and strip readout.

In situ measurement package

An in situ package, comprising a top hat plasma analyser (e.g. measuring protons and $\alpha$-particles) and a magnetometer, completes the instrument suite, making the mission self-sufficient, i.e. avoiding to rely on other spacecraft to monitor solar wind/magnetosheath plasma conditions. Moreover, simultaneous in situ measurements in both hemispheres (and on dawn and dusk flanks) allow the study of magnetospheric asymmetries. If one of the satellites is upstream of the quasi-parallel bow shock and the other is upstream of the quasi-perpendicular bow shock, we can compare characters of the plasma, waves and magnetic field parameters.

There is extensive heritage of in situ instrumentation on many previous and current space missions, e.g. Cluster and Cassini, THEMIS and MMS. Driver for the magnetometer is the determination of IMF discontinuities, and for the plasma analyser the detection of solar wind plasma (0.5 – 4.0 keV) variations at imager cadences with sufficient energy resolution to construct moments. Instrument requirements which follow are a resolution of 0.2 nT at a cadence of 1.5 s for the magnetometer and $\Delta E/E = 0.2$ at sub-second cadence for the plasma analyser. These are easily satisfied by the type of magnetometers and plasma analysers flying on Cluster and Cassini, THEMIS and MMS, and soon to fly on Solar Orbiter.

Additional observing facilities

A strong link with ground based measurement facilities, such as auroral imagers, radar arrays, magnetometer networks and global networks of ionospheric radars, will be essential in order to e.g. closely examine beading features detected from space; determine the onset of substorms, which will be then followed in detail with the space instrumentation; make timing measurements between a solar wind discontinuity reaching the subsolar magnetopause and the ground responding; provide measurements of convection and quantify reconnection rates. A space – ground facilities collaborative link of this kind has already been deployed and operated extensively in the context of the THEMIS mission, and is already planned for SMILE, so there are established precursors of the concepts described here.

Modeling support will also be required for the mission, in terms of both MHD and kinetic models. Again, this is an area under active development in preparation for the SMILE mission.

## 5. Which technology challenges would enable addressing the science questions proposed?

The proposed mission concept does not involve technological challenges in order to bear fruit: the instruments under consideration have already been flown or are at high TRL in view of forthcoming launches. A single spacecraft version of the concept mission outlined above has already been studied in detail in preparation of previous mission proposals. The two spacecraft mission configuration is estimated to be within the envelope of an ESA M-class mission on the basis of resource studies carried out by our NASA GSFC Core Proposing Team members.

It must be stressed that even in the unfortunate case of one spacecraft failing, the multi-faceted nature of the science goals described will produce breakthroughs in our understanding of solar wind – magnetosphere coupling, and its impact on geospace and the human domain. Strong synergy will exist with other space missions (e.g. at L1 and L5) and ground measurements likely to take place in the medium term to 2050. Moreover, since the spacecraft will spend most of the time in the solar wind, while often visiting the magnetosheath, magnetotail and the lobes, there will be the opportunity of carrying out a variety of other investigations of plasma conditions (e.g. turbulence, if in situ instrumentation with higher cadence could be part of the payload) making geospace an unparalleled plasma physics laboratory. Taking a larger scale holistic view, a mission such as the one envisaged here could represent the Earth-oriented component of an L-class mission incorporating elements linking solar, interplanetary, magnetospheric and ionospheric exploration, a truly global observatory of the heliosphere, with science aims perfectly aligned with those of crucial and still underdeveloped space weather research.

## Core Proposing Team

G. Branduardi-Raymont (Mullard Space Science Laboratory – University College London, UK)

M. Berthomier (Laboratoire de Physique des Plasmas, Paris, France)

Y. Bogdanova (Rutherford Appleton Laboratory, Didcot, UK)

J. C. Carter (University of Leicester, UK)

M. Collier (NASA Goddard Space Flight Center, USA)

A. Dimmock (Swedish Institute of Space Physics, Uppsala, Sweden)

M. Dunlop (Rutherford Appleton Laboratory, Didcot, UK)

R. Fear (University of Southampton, UK)

C. Forsyth (Mullard Space Science Laboratory – University College London, UK)

B. Hubert (University of Liege, Belgium)

E. Kronberg (University of Munich, Germany)

K. M. Laundal (University of Bergen, Norway)

M. Lester (University of Leicester, UK)

S. Milan (University of Leicester, UK)

K. Oksavik (University of Bergen, Norway)

N. Østgaard (University of Bergen, Norway)

M. Palmroth (University of Helsinki, Finland)

F. Plaschke (Space Research Institute, Austrian Academy of Sciences, Graz, Austria)

F. S. Porter (NASA Goddard Space Flight Center, USA)

I. J. Rae (Mullard Space Science Laboratory – University College London, UK)

A. Read (University of Leicester, UK)

A. Samsonov (Mullard Space Science Laboratory – University College London, UK)

S. Sembay (University of Leicester, UK)

Y. Shprits (German Research Centre for Geosciences, Potsdam, Germany)

D. G. Sibeck (NASA Goddard Space Flight Center, USA)

B. Walsh (Boston University, USA)

M. Yamauchi (Swedish Institute of Space Physics, Kiruna, Sweden)